\documentclass[%
 reprint,
 amsmath,amssymb,
 aps,twocolumn
]{revtex4-1}

\usepackage{graphicx}
\usepackage{dcolumn}
\usepackage{bm}
\usepackage[pagebackref=false,colorlinks,linkcolor=blue,citecolor=red]{hyperref}

\usepackage{tikz}
\usetikzlibrary{decorations.pathmorphing}
\usepackage{capt-of}
\usepackage{caption}
\usepackage{subcaption}

\usepackage{amssymb}
\usepackage{epsfig, float}

\usepackage{tikz}
\usetikzlibrary{decorations.pathmorphing}
\usepackage{capt-of}
\usepackage{caption}
\usepackage{subcaption}

\begin{document}

\preprint{APS/123-QED}

\title{Complexity Growth, Butterfly Velocity and Black hole Thermodynamics}

\author{Seyed Ali Hosseini Mansoori}
   \email{shossein@ipm.ir; shosseini@shahroodut.ac.ir}
   \affiliation{Faculty of Physics, Shahrood University of Technology, P.O.
Box 3619995161 Shahrood, Iran, \\ School of Astronomy,
Institute for Research in Fundamental Sciences (IPM), P.O. Box 19395-5531, Tehran, Iran}
  \author{Mohammad M. Qaemmaqami}
  \email{m.qaemmaqami@ipm.ir}
\affiliation{School of Particles and Accelerators,
Institute for Research in Fundamental Sciences (IPM),
P.O. Box 19395-5531, Tehran, Iran}



\begin{abstract}
	We propose a connection between the butterfly velocity and the complexity growth rate in the context of thermodynamics of black holes where the cosmological constant is interpreted as thermodynamic pressure. According to the Smarr formula of black hole systems,  there are different relations between butterfly velocity and complexity growth rate. The accuracy of this relationship has been checked for several models. \end{abstract}

\maketitle

\section{Introduction}
One of the recent holographic conjectures about the geometry of the  inside of black hole is that its growth is dual to the growth of quantum complexity \cite{Susskind:2014rva}. The definition of quantum complexity of a quantum state is the minimum number of simple gates for building a quantum circuit that constructs them from a reference state.

From AdS/CFT duality point of view, one of the famous conjectures is "complexity = volume" (CV), which is an example of the recent proposed connection between tensor network and geometry  \cite{Swingle:2009bg,Vidal:2007hda}. According to this conjecture, the volume of a maximal spacelike slice into the black hole interior, is proportional to the computational complexity of the dual CFT state \cite{Stanford:2014jda},
\begin{eqnarray}
\mathcal{C} (t_{L},t_{R})\sim \frac{V}{G_{N}l},
\end{eqnarray}
where $ V $ is the volume of the Einstein-Rosen bridge that joints two boundaries at the times $t_{L}$ and $t_{R}$ together, $ l $ is the AdS radius, and $ G_{N} $ is Newton's gravitational constant.
The other conjecture is "complexity = action" (CA) which implies that the quantum computational complexity of a holographic state is given by the on-shell action on the "Wheeler De-Witt" patch \cite{Brown:2015lvg},
\begin{eqnarray}
\mathcal{C}(\Sigma)= \frac{S_{WDW}}{\pi\hbar},
\end{eqnarray}
where $ \Sigma $ is the time slice which is the intersection of the asymptotic boundary and any Cauchy surface in the bulk. It has been shown in \cite{Brown:2015lvg} that the growth rate of quantum complexity will be bounded by 
\begin{eqnarray}
\frac{d\mathcal{C}}{dt}\leq \frac{2M}{\pi\hbar},
\end{eqnarray}
where $ M $ is the mass of black hole. For uncharged black hole the bound is usually saturated. Recently some works
have been done on quantum complexity and its different aspects in context of black holes and holography \cite{Alishahiha:2015rta,Couch:2016exn,Brown:2016wib,Brown:2017jil,Alishahiha:2017hwg,Bakhshaei:2017qud,Momeni:2016ira,Miao:2017quj,Qaemmaqami:2017lzs,Moosa:2017yvt}.\\

On the other hand, it was shown that chaos in thermal CFT may be described by
the propagation of the shock wave near horizon of an AdS black hole  \cite{Shenker:2013pqa,Roberts:2014isa,Leichenauer:2014nxa}, In the context of holography, the propagation of the shock
wave near the horizon provides a description of the butterfly effect in the dual field theory. Out-of-time order four point function between pairs of local operators diagnoses the butterfly effect in field theory side.
\begin{eqnarray}
\langle V_{x}(0)W_{y}(t)V_{x}(0)W_{y}(t)\rangle_{\beta},
\end{eqnarray}
where $ \beta $ is the inverse of the temperature. The butterfly effect may be seen by a sudden decay after the scrambling time $ t_{*} $ which is defined as $ t_{*}=\frac{\beta}{2\pi}log S $, where $ S $ is the entropy of black hole,
\begin{eqnarray}
\frac{\langle V_{x}(0)W_{y}(t)V_{x}(0)W_{y}(t)\rangle_{\beta}}{\langle V_{x}(0)V_{x}(0)\rangle_{\beta} \langle W_{y}(t)W_{y}(t)\rangle_{\beta}}\sim 1-e^{\lambda_{L}\big(t-t_{*}-\frac{|x-y|}{v_{B}}\big)},
\end{eqnarray}
where $ v_{B} $ is the butterfly velocity which characterizes speed at which the perturbation grows and $ \lambda_{L} $ is the Lyapunov exponent. The Lyapunov exponent is, $ \lambda_{L}={2\pi}/{\beta} $, where $ \beta $ is the inverse of the Hawking temperature. 
Recently some works have been done on butterfly effect and it's different aspects \cite{Alishahiha:2016cjk,Qaemmaqami:2017bdn,Qaemmaqami:2017jxz,Feng:2017wvc,Jahnke:2017iwi,Giataganas:2017koz}.
For example, in Ref. \cite{Feng:2017wvc}, authors found a universal formula for the butterfly velocities of planar black holes in the framework of Einstein's general relativity with respect to thermodynamical parameters such as the temperature, entropy and the volume conjugate to the pressure associated with the cosmological constant.\\

Moreover, the upper bound of the complexity growth rate is limited by the product of entropy and temperature. Therefore, using the Smarr formula relating the black hole mass to other thermodynamic quantities, we will be able to exploit a specific relation between the action growth and butterfly velocity through the thermodynamical parameters.\\ 

The structure of the paper is as follows. In next two Sections, we conjecture that there is the connection between complexity growth rate, the butterfly velocity, and thermodynamical parameters by paying attention to Smarr formulas. We check this conjecture for some systems such as BTZ black hole, neutral/charged black branes,  Lifshitz black brane, and the bulk solutions of a holographic model with momentum dissipation. 
In Section  \ref{Othersection}, we will discuss other evidence on the relationship between  the butterfly velocity and the complexity growth rate. Section \ref{Sec4} is devoted to the summary.

\section{Connection Between Complexity Growth Rate and Butterfly Velocity}\label{Sec2}


The relation between butterfly velocity $ v_{B} $ and thermodynamical variables is given by the following universal formula \cite{Feng:2017wvc} (See Appendix A):
\begin{eqnarray}\label{Eq1}
v_{B}^{2}= \frac{TS}{2PV},
\end{eqnarray}
where $ S $ is the entropy, $ T $ is the temperature, $ V $ is the thermodynamical volume, $ P $ is the pressure for a black hole. There is also a rough relation between complexity growth rate and the product of entropy and the temperature \cite{Stanford:2014jda,Brown:2015lvg} as follows,
\begin{eqnarray}\label{Eq2}
\frac{d\mathcal{C}}{dt}=\alpha TS.
\end{eqnarray}
It means that the complexity growth rate is proportional to $TS$ up to a numerical factor $\alpha$ \cite{Brown:2015lvg}. 
In \cite{Stanford:2014jda} the authors have shown that the complexity of a high-temperature thermofield double (TFD) state increases as
\begin{eqnarray}
\mathcal{C}(t_{L},t_{R})\propto TS|t_{L}+t_{R}|,
\end{eqnarray}
which is precisely the expected behavior of a quantum circuit model of complexity, i.e., the rate of computation measured in gates per unit time is proportional to the entropy times temperature \cite{Susskind:2014rva}. The entropy appears because it represents the width of the circuit and the temperature appears for the local interaction rate of the qubits \cite{Stanford:2014jda}.\\

One can see in three dimensions this relation Eq. (\ref{Eq2}) is exact at least for non-rotating BTZ black hole \cite{Banados:1992wn}. One can also test it for rotating BTZ black hole in 3D Einstein Gravity, New Massive Gravity (NMG) \cite{Bergshoeff:2009hq} and Minimal Massive 3D Gravity \cite{Bergshoeff:2014pca} which is a proposed model to resolve the bulk-boundary clash problem of Topologically Massive Gravity \cite{Deser:1981wh}.
For rotating BTZ black hole in 3D Einstein Gravity, the complexity growth rate is \cite{Brown:2015lvg}:
\begin{eqnarray}
\frac{d\mathcal{C}}{dt}= \frac{r_{+}^{2}-r_{-}^{2}}{4Gl^{2}}.
\end{eqnarray}
where $r_{-}$ and $r_{+}$ are the inner and outer horizon, respectively. One can also rewrite the above relation in terms of the inner and outer quantities as follows \cite{Cai:2016xho},
\begin{eqnarray}
\frac{d\mathcal{C}}{dt}=\frac{1}{2}\big(T_{+}S_{+}+T_{-}S_{-}\big),
\end{eqnarray}
where the temperatures and entropies defined on both horizons are $ T_{\pm}=\frac{r_{+}^{2}-r_{-}^{2}}{2\pi l^{2}r_{\pm}} $ and $ S_{\pm}=\frac{\pi r_{\pm}}{2G} $, respectively. For non-rotating BTZ black hole ($ r_{-}=0 $) we have
\begin{eqnarray}
\frac{d\mathcal{C}}{dt}= TS,
\end{eqnarray}
in which $ T= {r_{+}}/{2\pi l^{2}} $ and $ S= {\pi r_{+}}/{2G} $. It is obvious that the factor $\alpha=1$ for this case.
Now we can check the accuracy of the above relation for New Massive Gravity and Minimal Massive 3D Gravity. For New Massive Gravity the action growth is given by \cite{Alishahiha:2017hwg}:
\begin{eqnarray}
\frac{d\mathcal{C}}{dt}= \frac{r_{+}^{2}-r_{-}^{2}}{4Gl^{2}}\bigg(1-\frac{1}{2m^{2}l^{2}}\bigg),
\end{eqnarray}
for non-rotating case when the inner horizon goes to zero, we have
\begin{eqnarray}
\frac{d\mathcal{C}}{dt}= TS,
\end{eqnarray}
where \cite{Clement:2009gq},
\begin{eqnarray}
 S= \frac{\pi r_{+}}{2G}\big(1-\frac{1}{2m^{2}l^{2}}\big),\,\,\ ; \,\,\,\ T= \frac{r_{+}}{2\pi l^{2}}.
\end{eqnarray} 
 Moreover, in the case of the BTZ black hole in Minimal Massive 3D Gravity, the complexity growth rate is \cite{Qaemmaqami:2017lzs}:
\begin{eqnarray}
\frac{d\mathcal{C}}{dt}=\bigg[\sigma+\alpha\bigg(\frac{1-\alpha l^{2}\Lambda_{0}}{2\mu^{2} l^{2}(1+\sigma\alpha)^{2}}\bigg)\bigg]\frac{r_{+}^{2}-r_{-}^{2}}{4Gl^{2}},
\end{eqnarray}
where $ \sigma $, $ \alpha $ and $ \Lambda_{0} $ are the parameters of the model \cite{Bergshoeff:2014pca}. At the non-rotating limit, one can see $ {d\mathcal{C}}/{dt}= TS $, in which \cite{Setare:2015pva},
\begin{equation}
 S= \bigg[\sigma+\alpha\bigg(\frac{1-\alpha l^{2}\Lambda_{0}}{2\mu^{2} l^{2}(1+\sigma\alpha)^{2}}\bigg)\bigg] \frac{\pi r_{+}}{2G}, \,\,\,\ ; \,\,\,\ T= \frac{r_{+}}{2\pi l^{2}}.
\end{equation} 
If we replace the complexity growth rate instead of entropy times temperature $ TS $, in the the proposed equation for butterfly velocity Eq. (\ref{Eq1}), one can re-express the complexity growth rate in the terms of butterfly velocity and thermodynamical quantities as following relation. 
\begin{eqnarray}\label{eq2}
\frac{d\mathcal{C}}{dt}= 2\mathcal{N}_{1} v_{B}^{2}PV.
\end{eqnarray}
where $\mathcal{N}_{1}$ is usually a constant factor. In anistropic black hole, $\mathcal{N}_{1}$ may be a function of the outer horizon, $r_{+}$. For example, in Lifshitz black hole, because of the anisotropic scaling property between time and radial components of the Lifshitz metric, $\mathcal{N}_{1}$ is a function of $r_{+}$. By taking an alternate Smarr relation that has been employed for some planer black holes \cite{Brenna:2015pqa}, we can also state the correspondence between butterfly velocity and complexity action growth rate by the following general form,
\begin{equation}\label{aeq2}
\frac{d\mathcal{C}}{dt}= 2\mathcal{N}_{2} f(v_{B}^{2})PV.
\end{equation}
where $\mathcal{N}_{2}$ and $f$ are constant number and a polynomial function of $v_{B}^2$, respectively. 
Allow us now check the relation (\ref{eq2}) or (\ref{aeq2}) for some well-known cases
\subsection{Rotating/ non-rotating BTZ black hole}
For non-rotating BTZ black hole, the themodynamic variables and butterfly velocity are defined as \cite{Banados:1992wn,Shenker:2013pqa,Qaemmaqami:2017jxz} :
\begin{eqnarray}\label{eq3}
M&=& \frac{r_{+}^{2}}{8Gl^{2}} \;\;\ ; \;\;\ P=\frac{1}{8\pi Gl^{2}},\nonumber\\
V&=& \pi r_{+}^{2} \;\;\ ;\;\;\;\ v_{B}=1,
\end{eqnarray}
Substituting above relations of thermodynamic parameters and butterfly velocity in the Eq. (\ref{eq2}) for complexity growth rate, we have,
\begin{eqnarray}
\frac{d\mathcal{C}}{dt}= 2v_{B}^{2}PV= \frac{r_{+}^{2}}{4Gl^{2}},
\end{eqnarray}
which is in agreement with the previous result for non-rotating BTZ black hole \cite{Brown:2015lvg}.
For the rotating BTZ black hole case, one can obtain the following relation \cite{Couch:2016exn}:
\begin{eqnarray}
\frac{d\mathcal{C}}{dt}= 2v_{B}^2 P\big(V^{+}-V^{-}\big)=  \frac{r_{+}^{2}-r_{-}^{2}}{4Gl^{2}}, 
\end{eqnarray}
which satisfies the obtained result for rotating BTZ black hole \cite{Brown:2015lvg}. It is obvious that in above case $\mathcal{N}_{1}=1$.
\subsection{Neutral black brane}
As another example, let us consider a neutral black brane metric as
\begin{equation}
ds^2=-r^2 f(r)dt^2+\frac{1}{r^2 f(r)} dr^2+r^2 \delta_{ij} dx^i dx^j
\end{equation}
where $f(r)$ takes the below form
\begin{equation}
f(r)=1-\frac{ m}{r^{d+1}}
\end{equation}
The mass, entropy and temperature for such a black brane \cite{Brenna:2015pqa} are 
\begin{equation}
M=d\frac{ V_{d} m}{16 \pi G }=\frac{(d) V_{d}}{16 \pi G l^2} r_{+}^{d+1}; \hspace{0.25cm} S=\frac{V_{d}}{4G} r_{+}^{d}; \hspace{0.25cm} T=\frac{d+1}{4 \pi} \frac{r_{+}}{l^{2}}
\end{equation}
For this planar black hole, we can write down the Smarr relation by
\begin{equation}\label{sma1}
M=\frac{d}{d+1} TS
\end{equation}
It is also interesting that we can write an alternative Smarr relation for such case  \cite{Brenna:2015pqa}, namely 
\begin{equation}\label{sam2}
M=\frac{d}{d-1}TS-\frac{2}{d-1}PV
\end{equation}
where 
\begin{equation}\label{EE4}
P=\frac{d (d+1)}{16 \pi G l^2}, \hspace{0.15cm} V=\frac{V_{d}}{d+1} r_{+}^{d+1}
\end{equation}
Clearly, we see that Eq. (\ref{sma1}) is a special case of Eq. (\ref{sam2}) once it is recognized that $(d+1)PV=(d) TS$ for a neutral planar black hole.
According to Smarr relations in Eqs. (\ref{sma1}) and (\ref{sam2}),  complexity growth rate for such black branes at the late time \cite{Carmi:2017jqz} can be written as
\begin{equation}
\frac{dC}{dt}=2M=\Big(\frac{2 d}{d+1}\Big) TS
\end{equation}
or
\begin{equation}
\frac{dC}{dt}=2M=\frac{2d}{d-1}TS-\frac{4}{d-1}PV
\end{equation}
 By considering Eq. (\ref{VB1}), i.e. $TS=2 v_{B}^2 PV$, now we have
\begin{equation}
\frac{dC}{dt}=\Big(\frac{4d}{d+1}\Big) v_{B}^2 PV
\end{equation}
or
\begin{equation}\label{Schcom}
\frac{dC}{dt}=\frac{4d}{d-1} \Big(v_{B}^2-\frac{1}{d}\Big) PV
\end{equation}
which are consistent with Eqs. (\ref{eq2}) and (\ref{aeq2}). Therefore,  one can immediately read off
\begin{equation}
\mathcal{N}_{1}=\frac{2 d}{d+1}, \hspace{0.25cm} \mathcal{N}_{2}=\frac{2d}{d-1}, \hspace{0.25cm} f=v_{B}^2-\frac{1}{d}
\end{equation}
\subsection{Charged black brane}
The charged black planer solution is also given by
\begin{equation}
f(r)=1-\frac{m}{r^{d+1}}+\frac{Q^2}{r^{2d}}
\end{equation}
where $m$ and $Q$ are related to the mass and charge of the black brane solution, respectively. At the late time, the complexity action growth rate \cite{Alishahiha:2019cib} is 
\begin{equation}
\frac{dC}{dt}=\frac{V_{d}}{16 \pi G l^2} \Big[ f'(r_{+}) r_{+}^{d+2}-f'(r_{-}) r_{-}^{d+2}\Big]
\end{equation}
it may also be recast into the following form \cite{Alishahiha:2019cib}.
\begin{equation}
\frac{dC}{dt}=T_{+} S_{+}+T_{-}S_{-}
\end{equation}
where
\begin{equation}
T_{\pm}= \pm\frac{r_{\pm}^{2} f'(r_{\pm})}{4 \pi l^2} \hspace{0.5cm} S_{\pm}=\frac{V_{d}}{4 G} r_{\pm}^d
\end{equation}
By regarding the below thermodynamic variables \cite{Brenna:2015pqa},
\begin{equation}
V_{ \pm}=\frac{V_{d}}{d+1} r_{\pm}^{d+1}  \hspace{0.15cm} P=\frac{d (d+1)}{16 \pi G l^2}
\end{equation}
and using the below butterfly velocity,
\begin{equation}
v_{B \pm}^2=\pm r_{\pm} \frac{f'(r_{\pm})}{2 d} 
\end{equation}
we also rewrite the complexity growth rate as follows.
\begin{equation}
\frac{dC}{dt}=2P \Big[V_{+} v_{B+}^2-V_{-} v_{B-}^2\Big]
\end{equation}
Because of the bound on the butterfly velocity, i.e. $v_{B \pm} \leq v_{B}^{S}$ where $v_{B}^{S}$ is just the Schwarzschild value \cite{Mezei:2016zxg}, one can consider the upper bound on the complexity as follows
\begin{equation}
\frac{dC}{dt}\leq 2P (v_{B}^{S})^2 \Big[V_{+}-V_{-}\Big]
\end{equation}
\subsection{Lifshitz black brane}
Let us consider a general exact Lifshitz solution for Einstein-Dilaton-Maxwell theory presented in \cite{Tarrio:2011de} where the general solution is written
\begin{equation}
ds^2=-r^{2z} f(r) dt^2+\frac{dr^2}{r^2 f(r)}+r^2 \delta_{ij} dx^{i}dx^{j}
\end{equation}
with
\begin{equation}
f(r)=1-\frac{m}{r^{z+d}}
\end{equation}
Moreover, at late time the action growth rate for this case  \cite{An:2018xhv} is given by
\begin{equation}
\frac{dC}{dt}=2 \frac{(z+d-1)}{d}M=2 \frac{(z+d-1)}{z+d} TS
\end{equation}
Note that for $z>1$, we have $\frac{dC}{dt}\geq  2 M$ which indicates that the Lloyd's bound is violated at the late time.  
The mass, entropy and temperature 
of the Lifshitz black brane are 
\begin{equation}\label{EE3}
M=\frac{d V_{d}}{16 \pi G l^2} r_{+}^{d+z}, \hspace{0.15cm} S=\frac{V_{d}}{4 G } r_{+}^{d}, \hspace{0.15cm} T=\frac{(z+d)}{4 \pi l^2} r_{+}^{z}
\end{equation}
By using the below thermodynamic variables \cite{Brenna:2015pqa},
\begin{equation}\label{EE2}
P=\frac{(z+d) (z+d-1)}{16 \pi G l^2}, \hspace{0.15cm} V=\frac{d(z+1)V_{d}}{2 (d+z)(d+z-1)} r_{+}^{z+d}
\end{equation}
and butterfly velocity for $z>1$, 
\begin{equation}
v_{B}^2=\frac{z+d}{2d} r_{+}^{2z-2}
\end{equation}
Therefore, we arrive at 
\begin{equation}\label{EE1}
\frac{dC}{dt}=4 r_{+}^{2-2z} \frac{z+d-1}{z+d} v_{B}^2 P V 
\end{equation}
From Eq. (\ref{eq2}), one therefore reads off 
\begin{equation}
\mathcal{N}_{1}=2 r_{+}^{2-2z} \frac{z+d-1}{z+d}
\end{equation} 
where the coefficient $r_{+}^{2-2z}$ comes from the anisotropic scaling property between time and radial components of the metric. By regarding Eqs. (\ref{EE4}), (\ref{EE3}) and (\ref{EE2}), one can obtain the below bound on the complexity action growth that shows the violation of Lloyd's bound. 
\begin{equation}
\frac{dC}{dt}=4 \frac{(z+d-1)}{d+1} (v_{B}^{S})^2 PV \geq 2  (v_{B}^{S})^2 PV \geq 2M
\end{equation} 
   \section{Holographic model with momentum dissipation}\label{Hsection}
   In this section, allow us to check our conjecture for holographic models with momentum dissipation \cite{Andrade:2013gsa}. The simple bulk action for such systems can be written by
 \begin{eqnarray}\label{action1}
 S=\frac{1}{16 \pi G}\int d^4 x \sqrt{-g} \Big(R+\frac{6}{l^2}-\frac{1}{2} \sum_{i=1}^{2} \partial^{\mu}\phi_{i} \partial_{\mu} \phi_{i}\Big)\cr
\end{eqnarray}   
 where $\phi_{i}$ are massless scalar fields \cite{Andrade:2013gsa, Davison:2014lua}. The corresponding equations of motion are 
 \begin{eqnarray}
 G_{\mu \nu}-\frac{3}{L^2}g_{\mu\nu}&=& \frac{1}{2}\sum_{i=1}^{2}\Big[\partial_{\mu} \phi_{i} \partial_{\nu} \phi_{i}-2  g_{\mu \nu} \partial_{\lambda}\phi_{i}\partial^{\lambda}\phi_{i} \Big] \label{Einstien1}\cr
 \Box \phi_{i}&=&0 \,\,\ ; \,\,\,\ i=1,2
 \end{eqnarray}
By taking the following ansatz for the metric and the scalar fields,
\begin{eqnarray}
\nonumber ds^2=-r^2 f(r) dr^2&+&\frac{1}{r^2 f(r)} dr^2+r^2 \delta_{ij} dx^{i} dx^{j}\\
 \phi_{i}&=&m \delta_{ij}x^{j} \hspace{0.5cm} i,j=1,2
\end{eqnarray}
and plugging them into Eq. (\ref{Einstien1}), one can obtain a black brane solution as
\begin{equation}
f(r)=\frac{1}{l^2}-\frac{m^2}{2 r^2}-\frac{r_{+}^3}{r^3} \Big(\frac{1}{l^2 }-\frac{m^2}{2 r_{+}^2}\Big)
\end{equation}
where $r_{+}$ is the outer radius. Moreover, the temperature, mass, and entropy are given by \cite{Andrade:2013gsa}
\begin{eqnarray}
T&=&\frac{1}{4\pi}\frac{d}{dr}\Big(r^2 f(r)\Big)|_{r=r_{+}}=\frac{r_{+}}{4 \pi} \Big[\frac{3}{l^2}-\frac{m^2}{2 r_{+}^2}\Big]\\
M&=&\frac{ V_{2}}{8 \pi G l^2} r_{+}^3 \Big(1-\frac{m^2 l^2}{2 r_{+}^2}\Big)\hspace{0.5cm} S= \frac{V_{2}}{4 G} r_{+}^2 
\end{eqnarray}
Note that we only have real positive values for temperature in the range where $m$ restricts to $0 \le m < \frac{\sqrt{6} r_{+}}{L}$. Moreover, if we choose the $m$ parameter in the range  $0 \le m< \sqrt{\frac{3}{2}} \frac{r_{+}}{L}$ ( It comes from the roots of $f(r)=0$.), the black brane has only one horizon and the inner horizon is absent. In other words, when $0 \le m< \sqrt{\frac{3}{2}} \frac{r_{+}}{L}$, the Penrose diagram of this black brane is similar to the one for the AdS Schwarzschild black brane \cite{Cai:2017sjv}. 

\begin{figure}[h]
\begin{center}
\includegraphics[width=.50\textwidth]{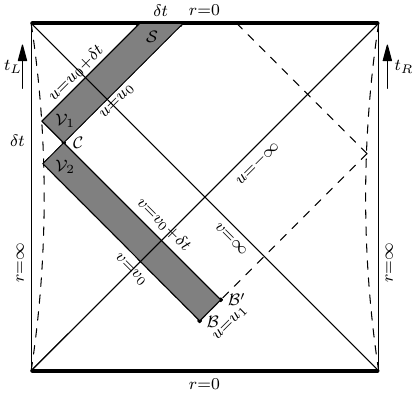}
\end{center}
\caption{The Wheeler-DeWitt (WDW) patch of black holes with single horizon. Figure has been taken from \cite{Meng:2018vtl}.}
\label{fig1}
\end{figure}

Now we are interested in calculating the action growth rate of the Wheeler-DeWitt (WDW) patch in our case with single horizon, as shown in Fig. \ref{fig1}, where the origin $r = 0$ is a spacelike singularity. 

 Here, we use the method proposed in \citep{Lehner:2016vdi, Cai:2017sjv} in order to obtain the complexity action growth rate. To proceed our calculations, it is convenient to define the null coordinates, 
\begin{equation}
du=dt+\frac{1}{r^2 f(r)}dr, \hspace{1cm} dv=dt-\frac{1}{r^2 f(r)}dz
\end{equation}  
and
\begin{equation}
u=t+r^{*}(r),\hspace{0.25cm} v=t-r^{*}(r),\hspace{0.25 cm} r^{*}=\int \frac{1}{ r^2 f(r)}dr
\end{equation} 
under the null coordinates the metric is written as
\begin{equation}
ds^2=-r^2 f(r) du^2+2 drdu+r^2 \delta_{ij} dx^{i} dx^{j}
\end{equation}
or
\begin{equation}
ds^2=-r^2 f(r) dv^2-2 dvdr+r^2 \delta_{ij} dx^{i} dx^{j}
\end{equation}
The action difference $\delta \mathcal{I}=\mathcal{I}(t_{0}+\delta t)-\mathcal{I}(t_{0})$ is a sum of the volume contributions from the regions $\mathcal{V}_{1}$ and $\mathcal{V}_{2}$, the surface contribution from the spacelike segment $\mathcal{S}$, and the joint contributions from the surfaces at $\mathcal{B}'$ and $\mathcal{B}$, i.e.,
\begin{eqnarray}
\nonumber \delta {\cal I}& =& \mathcal{I}_{\mathcal{V}_{1}}-\mathcal{I}_{\mathcal{V}_{2}}-\frac{1}{8 \pi G} \int_{\mathcal{S}}K d\Sigma\\
&+& \frac{1}{8 \pi G} \oint_{\mathcal{B}'}a dS-\frac{1}{8 \pi G} \oint_{\mathcal{B}}a dS
\end{eqnarray}
where $d \Sigma$ is the volume element on $\mathcal{S}$, $dS$ is the surface element on $\mathcal{B}$ and $\mathcal{B}'$. Moreover, $K$ is trace of the extrinsic curvature and $a$ will be illustrated later.

As shown in the Fig \ref{fig1}, the past and future null boundaries on the left of
the first WDW patch are labeled by $u=u_{0}$ and $v=v_{0}$, respectively. After a time shift $\delta t$, these null boundaries become $u=u_{0}+\delta t$ and $v=v_{0}+\delta t$. 
The region $\mathcal{V}_{1}$ is also surrounded
by the null surfaces $u=u_{0}$, $u=u_{0}+\delta t$, $v=v_{0}+\delta t$, and the spacelike surface $\mathcal{S}$ at $r=\epsilon$ near the singularity $r=0$. At the end of calculations, we will take the limit
$\epsilon \to 0$. Moreover, the surface $v=v_{0}+\delta t$ is described by $r = \rho(u)$, where $\rho(u)$ is defined by
$r^{*}(\rho)=-\frac{1}{2}(v_{0}+\delta t-u)$. 
 
Making use of Eq. (\ref{Einstien1}), the bulk action of region $\mathcal{V}_{1}$ is
\begin{equation}\label{actionv1}
\mathcal{I}_{\mathcal{V}_{1}}=-\frac{3V_{2}}{8 \pi G l^2} \int_{u_{0}}^{u_{0}+\delta t}du \int^{\rho(u)}_{\epsilon} dr r^2
\end{equation}
 The volume integral in $\mathcal{V}_{2}$ should be calculated
in the $(v,r)$ coordinates, where the surfaces $u=u_{0,1}$  are described by $r=\rho_{0,1}(v)$ which satisfies
$r^{*}(\rho_{0,1})=-\frac{1}{2}(v-u_{0,1})$. Therefore, the contribution of $\mathcal{V}_{2}$ to the action is
\begin{equation}
\mathcal{I}_{\mathcal{V}_{2}}=-\frac{3V_{2}}{8 \pi G l^2} \int_{v_{0}}^{v_{0}+\delta t}dv \int_{\rho_{1}(v)}^{\rho_{0}(v)} dr r^2
\end{equation}   
  By changing the variable $u \to u_{0}+v_{0}+\delta-v$ in Eq. (\ref{actionv1}), the integration bound $[u_{0},u_{0}+\delta t]$ becomes $[v_{0}+\delta t,v_{0}]$ and $\rho(u)$ which satisfies $r^{*}(\rho)=-\frac{1}{2} (v_{0}+\delta t-u) \to -\frac{1}{2} (v-u_{0}) $ becomes $\rho_{0}(v)$. At the late time, $\rho_{1}(v)\to r_{+}$, we have    
\begin{eqnarray}
\mathcal{I}_{{\mathcal{V}}_{1}}-\mathcal{I}_{{\mathcal{V}}_{2}}&=& -\frac{3V_{2}}{8 \pi G l^2} \int_{v_{0}}^{v_{0}+\delta t} dv \int^{r_{+}}_{\epsilon} dr r^2\\
\nonumber &=& -\frac{3V_{2}}{8 \pi G l^2}  \delta t \int^{r_{+}}_{0} dr r^2=-\frac{r_{+}^3}{8 \pi G l^2} \delta t 
\end{eqnarray}
   Next, we consider the contribution from the spacelike surface $\mathcal{S}$ at $r=\epsilon$. The trace of the extrinsic curvature for a constant-$r$ hypersurface with the outward unit normal vector, $n^{\mu}=r \sqrt{f(r)} \delta^{\mu}_{r}$, is obtained as
\begin{equation}\label{Extrinsic}
K=\nabla_{\mu}n^{\mu}=\frac{1}{r^2 } \frac{d}{dr}\Big(r^3 \sqrt{f(r)}\Big)
\end{equation} 
By using above equation, we get 
\begin{eqnarray}
\mathcal{I}_{\mathcal{S}}&=&-\frac{1}{8 \pi G} \int_{\mathcal{S}} K d \Sigma=-\frac{1}{8 \pi G} \int_{\mathcal{S}}d^3 x  K \sqrt{-\gamma} \\
\nonumber &=& -\frac{V_{2}}{16 \pi G} \delta t \Big(6 r^3 f(r)+r^4 f'(r)\Big)|_{r \to \epsilon}
\end{eqnarray}
 in which $\gamma$ is the induced metric on the boundary. Note that the above term will be vanished as $\epsilon$ goes to zero in our case. Finally we calculate the contributions from the joints $\mathcal{B}'$ and $\mathcal{B}$. The function $a$ is defined by
\begin{equation}
a=\ln\Big(-\frac{1}{2} k_{1}.k_{2}\Big) 
\end{equation}
 where $k_{1}$ is the null normal to the hypersurface $v=const$ and $k_{2}$ is the null normal to the hypersurface $u=const$. Thus
   \begin{equation}
    k_{1}=-n_{1} \partial_{v}, \hspace{0.25cm} k_{2}=n_{2} \partial_{ u}=n_{2} \Big(\partial_{v}+\frac{2}{r^2 f(r)} \partial_{r}\Big)
   \end{equation}
where $n_{1}$ and $n_{2}$ are positive constants which are chosen to satisfy the asymptotic normalizations, $k_{1}.t_{L}=n_{1}$ and $k_{2}.t_{R}=-n_{2}$. With the choice, we have $a=-\ln(-{f(r)}/{n_{1}n_{2}})$, and 
\begin{eqnarray}
\mathcal{I}_{\mathcal{B} \mathcal{B}'}&=& \frac{1}{8 \pi G} \oint_{\mathcal{B}'} a dS-\frac{1}{8 \pi G} \oint_{\mathcal{B}} a dS\\
\nonumber &=& \frac{V_{2}}{8 \pi G}\Big( \mathcal{H}(r_{\mathcal{B}'})-\mathcal{H}(r_{\mathcal{B}})\Big)
\end{eqnarray}
in which $\mathcal{H}=r^2 a$. By making a Taylor expansion of $\mathcal{H}(r)$ around $r=r_{\mathcal{B}}$ and using $r_{\mathcal{B}'}-r_{\mathcal{B}}=-\frac{1}{2}r^2 f(r) \delta t$, we have
\begin{equation}
\mathcal{I}_{\mathcal{B} \mathcal{B}'}= \frac{V_{2}}{8 \pi G}\delta t  r^2 f(r) \Big[\frac{r^2 f'(r)}{ f(r)}-2 r\Big(1+ \ln\big(\frac{-f(r) r^2}{n_{1} n_{2}}\Big)\Big)\Big]|_{r=r_{\mathcal{B}}}
\end{equation}
At late times, $r_{\mathcal{B}}$ approaches $r_{+}$, therefore we have,
\begin{equation}
\mathcal{I}_{\mathcal{B} \mathcal{B}'}= \frac{V_{2}}{8 \pi G}\delta t \Big(r^4 f'(r)\Big)|_{r=r_{+}}=\frac{r_{+}^3 V_{2} \delta t}{8 \pi G l^2} \Big(3-\frac{ l^2 m^2}{r_{+}^2}\big)
\end{equation}
Collecting our calculations, adding up all the contributions to $\delta \mathcal{I}$, we finally arrive at
\begin{equation}\label{action growth 3}
\frac{d C}{d t}=\frac{\delta \mathcal{I}}{\delta t}= 2 M
\end{equation}
It means that Lloyd's bound is saturated in our case.  In addition, our case satisfies the Smarr formula,
\begin{equation}\label{smarr1}
M=2 TS-2 P V
\end{equation}
along with the first thermodynamic law
\begin{equation}
dM=TdS+VdP
\end{equation}
where
\begin{equation}
V=\frac{V_{d}}{3} r_{+}^3, \hspace{1cm} P=\frac{3}{8 \pi G l^2}
\end{equation}
Considering Eqs. (\ref{action growth 3}), (\ref{smarr1}), and (\ref{VB1}), we therefore have
\begin{equation}
\frac{dC}{dt}=8 P V \Big( v_{B}^2-\frac{1}{2}\Big)
\end{equation}
which is in agreement with our conjecture, Eq. (\ref{aeq2}) when one takes $f=v_{B}^2-\frac{1}{2}$ and $\mathcal{N}_{2}=4$. In the absence of scalar fields ($m \to 0$), this identity shifts to Eq. (\ref{Schcom}) in four dimensions.  Notice that Eq. (\ref{eq2}) cannot be correct here, because  we are not able to obtain any direct relation between $PV$ and $TS$.
\section{Other evidence} \label{Othersection}
   
   Here we consider some evidence that show the relation between butterfly velocity and complexity growth rate.  
   
   In Ref. \cite{Alishahiha:2016cjk} has been shown that the presence of higher derivative terms in the action of the critical gravity leads to propagate the shock wave with two different butterfly velocities  which are given by
\begin{eqnarray}
\nonumber v_{B}^{(1)}&=&\sqrt{\frac{D-1}{2(D-2)}},\\  v_{B}^{(2)}&=&\sqrt{\frac{D-1}{2(D-2)}}\left(1+\frac{2l^{2}}{(D-1)(D-2)M^{2}}\right)^{-\frac{1}{2}}
\end{eqnarray}
where $ M $ is the mass of massive spin-2 mode and $v_{B}^{(1)}$ is the butterfly velocity of Einstein Gravity in D-dimension \cite{Shenker:2013pqa}. From these expression it is clear that $ v_{B}^{(1)}> v_{B}^{(2)} $, it means that adding higher derivative terms to Einstein gravity action causes to decrease the butterfly velocity. A similar effect has also been reported by one of the authors in Ref. \cite{Qaemmaqami:2017bdn}.  More precisely, the butterfly velocity at the critical point in third order Lovelock Gravity in $ D=7 $ is less than the butterfly velocity at the critical point in Einstein-Gauss-Bonnet Gravity in $ D=7 $ which is less than the butterfly velocity in D = 7 in Einstein Gravity, $  v_{B}^{E.H}>v_{B}^{E.G.B}>v_{B}^{3rd\,\,Lovelock}  $ \cite{Qaemmaqami:2017bdn}.  On the other hand, the effects of higher derivatives terms in the critical gravity perturbed by a shock wave slows down the rate of complexity growth \cite{Alishahiha:2017hwg}.  Therefore we can argue that it's an evidence of connection between butterfly velocity and complexity growth rate.\\


Moreover, there is a correspondence between butterfly velocities and the central charges of the dual conformal field theory \cite{Qaemmaqami:2017jxz}. The central charges of the dual 2D CFT of Topologically Massive Gravity(TMG) reads \cite{Li:2008dq}:
\begin{eqnarray}
c_{L}=\frac{3l}{2G}\Big(1-\frac{1}{\mu l}\Big) \,\,\ ; \,\,\      c_{R}=\frac{3l}{2G}\Big(1+\frac{1}{\mu l}\Big).     
\end{eqnarray}
Clearly,  at two critical points of the Topologically Massive Gravity (TMG), $ \mu l=1 $ and $ \mu l=-1 $ there are two different chiral modes, right-moving and left-moving respectively,
\begin{eqnarray}\label{eq4}
\mu l=1, \,\,\ ; \,\,\   c_{L}=0,  \,\,\ ; \,\,\            c_{R}=\frac{3l}{G},\cr\nonumber\\
\mu l=-1, \,\,\ ; \,\,\  c_{L}=\frac{3l}{G},  \,\,\ ; \,\,\            c_{R}=0.
\end{eqnarray}
Moreover, at the critical points of TMG, the butterfly velocities yield as follows \cite{Qaemmaqami:2017jxz}:
\begin{eqnarray}
\mu l&=&1,        \,\,\ ; \,\,\    v_{B}^{(3)}=0,        \,\,\ ; \,\,\            v_{B}^{(2)}=1,\nonumber\\
\mu l&=&-1,   \,\ ; \,\  v_{B}^{(3)}=-1,        \,\,\ ;\,\,\    v_{B}^{(2)}=0,
\end{eqnarray}
It means that the theory is chiral at the critical points, $ \mu l=1 $ and $ \mu l=-1 $. These relations are similar to the relations for the central charges of the dual 2D conformal field theory Eq. (\ref{eq4}) at the critical points where the theory is chiral. 
On the other hand, the complexity is also proportional to the number of degrees of freedom or central charges \cite{Brown:2015lvg,Alishahiha:2015rta,Couch:2016exn}. Particularly for a 2D CFT when we consider a subsystem, $ A $ with the length $ L $, the complexity of the subsystem is \cite{Alishahiha:2015rta}:
\begin{eqnarray}
\mathcal{C}_{A}=\frac{cL}{12\pi\varepsilon}-\frac{c}{24},
\end{eqnarray}
where $ \varepsilon $ is the cutoff length of the field theory. Therefore, we can get the other evidence for connection between butterfly velocity and complexity growth rate. 

\section{Summary}\label{Sec4}
In this paper, we studied the connection between butterfly velocity and complexity growth rate by thermodynamical parameters according to Smarr relations and the recent proposed universal formula for butterfly velocity with respect to thermodynamic variables \cite{Feng:2017wvc}.  More precisely,
for some black holes whose mass is proportional to the product of the entropy and temperature ($TS$) by Smarr formula, $M= aTS$ in which $a$ is constant number, the complexity growth rate is related to butterfly velocity by Eq. (\ref{eq2}). While for the other cases with general Smarr relation $M=a'TS+b'PV$ where $a'$ and $b'$ are constant,  this relationship expresses by Eq. (\ref{aeq2}). It should be noted that when one can exploit another Smarr relation like $TS=c'PV$ ($ c'$ is constant),  both conjectures manifest the same result. 

Moreover, we brought several arguments based on higher order correction and central charge which show a new possible connection between the butterfly velocity and complexity growth rate.
 
It is also interesting to study the relationship between the complexity growth and butterfly velocity with entanglement spreading and entanglement velocity. Moreover, the study of the connection between complexity, chaos and tensor networks in the context of holography and black hole physics might be fascinating.

\section*{Acknowledgment}
We would like to thank Navid Abbasi and Ali Naseh for useful discussions and also Farid Charmchi, Dariush Kaviani and Siavash Neshatpour for comments on the manuscript. We would like to thank the referee for his/her instructive comments.


\appendix
\section{Butterfly velocity}
 Here we first study a shock wave solution of a generic static AdS black brane in $d+2$ dimensions when their solution is perturbed by injecting a small amount of energy \cite{Shenker:2013pqa,Roberts:2014isa,Leichenauer:2014nxa}. In similar to Ref. \cite{Feng:2017wvc}, we then try to write the square of the butterfly velocity as a function of thermodynamic variables like temperature, entropy and thermodynamic valume. Let us consider a metric form for these black branes as
 \begin{eqnarray}
ds^2&=& -h(r)  dt^2+{1\over f(r)}dr^2 +{r^2} dx^{i} dx^{j} \,,
\label{planar metric ansatz}
\end{eqnarray}
Now it is convenient to re-write the black brane solution in the Kruskal coordinates:
\begin{eqnarray}\label{d}
uv=-e^{\sqrt{f'(r_h)h'(r_h)}r_*},  u/v=-e^{-\sqrt{f'(r_h)h'(r_h)}t}
\end{eqnarray}
where $dr_*=\frac{dr}{\sqrt{f(r)h(r)}}$. In this coordinate system, the metric can be recast into the following form
\begin{eqnarray}\label{met1}
 \nonumber ds^2\,&=&\,2\,A(uv)\,du\,dv\,+\,B(uv)\,dx^i\,dx^i,
\end{eqnarray}
 where functions appearing in the metric are written as follows.
\begin{eqnarray}\label{e}
A(uv)\,=\,\,\frac{2}{u\,v}\,\frac{h(r)}{f'(r_h)\,h'(r_h)}, \ \ B(uv)\,=\,r^2
\end{eqnarray}
Moreover, the horizon location $r=r_h$ is mapped into $uv=0$. Now let us perturb this metric by releasing a particle from the boundary at $x=0$. At the late time $t_{w}>\beta$
the localized stress tensor is given by
\begin{equation}
\delta T_{uu}^{shock}\,=\,E\,e^{2\,\pi\,t_w/\beta}\,\delta(u)\,\delta(x)
\end{equation}
where $\beta=1/T$ and $E$ is the asymptotic energy of the particle. Moreover, the back-reaction of this pulse of energy in the left side of the geometry is obtained when there is a shift $v\rightarrow v\,+\,h(x,t_w)$, where $h(x,t_w)$ is a function which can be obtained from Einstein's equations. Therefore, the metric is perturbed as,
\begin{eqnarray}\label{met2}
ds^2\,&=&\,2\,A(uv)\,du\,dv\,+\,B(uv)\,dx^i\,dx^i\,\\
\nonumber &-&\,2\,A(uv)\,h(x,t_w)\,\delta(u)\,du^2
\end{eqnarray}
 Note that the function $h(x,t_w)$ can be found form the perturbed Einstein's equations,
\begin{equation}
G_{\mu\nu}+\delta G_{\mu\nu}=T_{\mu\nu}+\delta T_{\mu\nu}+\delta T_{\mu\nu}^{shock}
\end{equation}
Notice that the Einstein's equations $G_{\mu\nu}$ are satisfied in background metric (\ref{met1}), one can derive a second order differential equation for $h(x,t_w)$ near the horizon at the leading order of the perturbation by substituting the perturbed ansatz metric (\ref{met2}) in perturbed Einstein equation as
\begin{eqnarray}\label{c4}
\left[\partial_i^2-m^2\right]h(x^i,t_w)=\frac{ B(0)E e^{2\pi t_w/\beta}\delta(x)}{
   A(0)}
\end{eqnarray}
where the effective mass reads:
\begin{eqnarray}\label{c5}
m^2=\frac{(d)
    B'(0)}{
   A(0)}
\end{eqnarray}
 At large distances $|x|$, the solution takes the following form:
\begin{equation}\label{reH}
h(x,t_w)\,\sim\,\frac{E\,e^{\frac{2\,\pi}{\beta}(t_w\,-\,t^*)\,-\,m\,|x|}}{|x|^{\frac{d-1}{2}}}
\end{equation}
where $t^*$ is the scrambling time. Now we can read off  the Lyapunov exponent and the butterfly velocity as $\lambda_L\,=\,\frac{2\,\pi}{\beta}$, and  $v_B\,=\,\frac{2\,\pi}{\beta\,m}$, respectively. Finally in order to obtain the butterfly velocity, we need to rewrite the function $A(0)$, $B(0)$ and their derivatives in the original $(t,r,x,y)$ coordinates. Near the horizon,
\begin{eqnarray}
f(r)&\approx & f'(r_{h}) (r-r_{h})+\frac{f''(r_{h})(r-r_{h})^2}{2}+...\\
\nonumber h(r)&\approx & h'(r_{h}) (r-r_{h})+\frac{h''(r_{h})(r-r_{h})^2}{2}...
\end{eqnarray}
we also have,
\begin{equation}
r^*=\int \frac{1}{\sqrt{h(r)f(r)}} \approx \frac{1}{\sqrt{f'(r_{h})h'(r_{h})}} ln (r-r_{h})
\end{equation}
Therefore, above relations imply that
\begin{eqnarray}
uv &\approx & -c_{0} (r-rh)\\
A(0) &\approx & -\frac{4}{c_{0}f'(r_{h})}\\
\nonumber A'(0)&=&\frac{dA(uv)}{d(uv)}|_{u=0}=\frac{dA(uv)}{dr^*} \frac{dr^*}{d(uv)}|_{r_{h}}\\
 &=&\frac{h''(r_{h})}{c_{0}^{2} f'(r_{h}) h'(r_{h})}\\
&\,\,\,\,\,\,\,\ &  B'(0)=-\frac{2 r_{h}}{c_{0}}
\end{eqnarray}
Considering above equations and using Einstein equation, the effective mass formulas reduces to
\begin{equation}
m^2=(d) r_{h} f'(r_{h})
\end{equation}
One can now obtain the butterfly velocity as follows:
\begin{equation}
v_B^2\,=\sqrt{\frac{h'(r_{h})}{f'(r_{h})}}\frac{2 \pi\, l^2 T\,}{ r_h d}
 \label{butterfly velocity}
\end{equation}
By using the thermodynamic variables for such black branes, one can rewrite the butterfly velocity by the following relation \cite{Feng:2017wvc}.
\begin{equation}\label{VB1}
v_{B}^{2}= \frac{TS}{2PV}.
\end{equation} 
where for the neutral black brane in ($d+2$) dimensions, the entropy $S$, temperature $T$, the thermodynamic volume $V$, and the pressure, $ P $ are defined by
\begin{eqnarray}
T=\frac{\sqrt{h'(r_{+}) f'(r_{+})}}{4 \pi l^{2}} \hspace{0.5cm} S=\frac{r_{+}^{d} V_{d}}{4 G} \\
V=\frac{r_{+}^{d+1} V_{d}}{d+1} \hspace{0.5cm} P=\frac{d(d+1)}{16 \pi G l^2}
\end{eqnarray}
where $V_{d}$ is the surface area of the space orthogonal to fixed $(t,r)$ surfaces.



\begin{thebibliography}{99}

\bibitem{Susskind:2014rva} 
L.~Susskind,
``Computational Complexity and Black Hole Horizons,''
[Fortsch.\ Phys.\  {\bf 64}, 24 (2016)]
Addendum: Fortsch.\ Phys.\  {\bf 64}, 44 (2016)
[arXiv:1403.5695 [hep-th], arXiv:1402.5674 [hep-th]].



\bibitem{Swingle:2009bg} 
B.~Swingle,
``Entanglement Renormalization and Holography,''
Phys.\ Rev.\ D {\bf 86}, 065007 (2012)
[arXiv:0905.1317 [cond-mat.str-el]].


\bibitem{Vidal:2007hda} 
G.~Vidal,
``Entanglement Renormalization,''
Phys.\ Rev.\ Lett.\  {\bf 99}, no. 22, 220405 (2007)
[cond-mat/0512165].


\bibitem{Stanford:2014jda} 
D.~Stanford and L.~Susskind,
``Complexity and Shock Wave Geometries,''
Phys.\ Rev.\ D {\bf 90}, no. 12, 126007 (2014)
[arXiv:1406.2678 [hep-th]].

\bibitem{Brown:2015lvg} 
A.~R.~Brown, D.~A.~Roberts, L.~Susskind, B.~Swingle and Y.~Zhao,
``Complexity, action, and black holes,''
Phys.\ Rev.\ D {\bf 93}, no. 8, 086006 (2016)
[arXiv:1512.04993 [hep-th]].



\bibitem{Alishahiha:2015rta} 
M.~Alishahiha,
``Holographic Complexity,''
Phys.\ Rev.\ D {\bf 92}, no. 12, 126009 (2015)
[arXiv:1509.06614 [hep-th]].


\bibitem{Couch:2016exn} 
J.~Couch, W.~Fischler and P.~H.~Nguyen,
``Noether charge, black hole volume, and complexity,''
JHEP {\bf 1703}, 119 (2017)
[arXiv:1610.02038 [hep-th]].


\bibitem{Brown:2016wib} 
A.~R.~Brown, L.~Susskind and Y.~Zhao,
``Quantum Complexity and Negative Curvature,''
Phys.\ Rev.\ D {\bf 95}, no. 4, 045010 (2017)
[arXiv:1608.02612 [hep-th]].



\bibitem{Brown:2017jil} 
  A.~R.~Brown and L.~Susskind,
  ``Second law of quantum complexity,''
  Phys.\ Rev.\ D {\bf 97}, no. 8, 086015 (2018)
  [arXiv:1701.01107 [hep-th]].



\bibitem{Alishahiha:2017hwg} 
M.~Alishahiha, A.~Faraji Astaneh, A.~Naseh and M.~H.~Vahidinia,
``On complexity for F(R) and critical gravity,''
JHEP {\bf 1705}, 009 (2017)
[arXiv:1702.06796 [hep-th]].


\bibitem{Bakhshaei:2017qud} 
E.~Bakhshaei, A.~Mollabashi and A.~Shirzad,
``Holographic Subregion Complexity for Singular Surfaces,''
Eur.\ Phys.\ J.\ C {\bf 77}, no. 10, 665 (2017)
[arXiv:1703.03469 [hep-th]].


\bibitem{Momeni:2016ira} 
D.~Momeni, M.~Faizal, S.~Bahamonde and R.~Myrzakulov,
``Holographic complexity for time-dependent backgrounds,''
Phys.\ Lett.\ B {\bf 762}, 276 (2016)
[arXiv:1610.01542 [hep-th]].


\bibitem{Miao:2017quj} 
  Y.~G.~Miao and L.~Zhao,
  ``Complexity-action duality of the shock wave geometry in a massive gravity theory,''
  Phys.\ Rev.\ D {\bf 97}, no. 2, 024035 (2018)
  [arXiv:1708.01779 [hep-th]].



\bibitem{Qaemmaqami:2017lzs} 
M.~M.~Qaemmaqami,
``Complexity growth in minimal massive 3D gravity,''
Phys.\ Rev.\ D {\bf 97}, no. 2, 026006 (2018)
[arXiv:1709.05894 [hep-th]].



\bibitem{Moosa:2017yvt} 
  M.~Moosa,
  ``Evolution of Complexity Following a Global Quench,''
  JHEP {\bf 1803}, 031 (2018)
  [arXiv:1711.02668 [hep-th]].

	
	
\bibitem{Shenker:2013pqa} 
S.~H.~Shenker and D.~Stanford,
``Black holes and the butterfly effect,''
JHEP {\bf 1403}, 067 (2014)
[arXiv:1306.0622 [hep-th]].



\bibitem{Roberts:2014isa} 
D.~A.~Roberts, D.~Stanford and L.~Susskind,
``Localized shocks,''
JHEP {\bf 1503}, 051 (2015)
[arXiv:1409.8180 [hep-th]].



\bibitem{Leichenauer:2014nxa} 
S.~Leichenauer,
``Disrupting Entanglement of Black Holes,''
Phys.\ Rev.\ D {\bf 90}, no. 4, 046009 (2014)
[arXiv:1405.7365 [hep-th]].


\bibitem{Alishahiha:2016cjk} 
M.~Alishahiha, A.~Davody, A.~Naseh and S.~F.~Taghavi,
``On Butterfly effect in Higher Derivative Gravities,''
JHEP {\bf 1611}, 032 (2016)
[arXiv:1610.02890 [hep-th]].


\bibitem{Qaemmaqami:2017bdn} 
M.~M.~Qaemmaqami,
``Criticality in third order lovelock gravity and butterfly effect,''
Eur.\ Phys.\ J.\ C {\bf 78}, no. 1, 47 (2018)
[arXiv:1705.05235 [hep-th]].


\bibitem{Qaemmaqami:2017jxz} 
M.~M.~Qaemmaqami,
"Butterfly Effect in 3D Gravity,''
Phys.\ Rev.\ D {\bf 96}, 106012 (2017)
[arXiv:1707.00509 [hep-th]].


\bibitem{Feng:2017wvc} 
X.~H.~Feng and H.~Lu,
``Butterfly Velocity Bound and Reverse Isoperimetric Inequality,''
Phys.\ Rev.\ D {\bf 95}, no. 6, 066001 (2017)
[arXiv:1701.05204 [hep-th]].



\bibitem{Jahnke:2017iwi} 
  V.~Jahnke,
  ``Delocalizing entanglement of anisotropic black branes,''
  JHEP {\bf 1801}, 102 (2018)
  [arXiv:1708.07243 [hep-th]].


\bibitem{Giataganas:2017koz} 
  D.~Giataganas, U.~Gürsoy and J.~F.~Pedraza,
  ``Strongly-coupled anisotropic gauge theories and holography,''
  Phys.\ Rev.\ Lett.\  {\bf 121}, no. 12, 121601 (2018)
  [arXiv:1708.05691 [hep-th]].

\bibitem{Deser:1981wh} 
S.~Deser, R.~Jackiw and S.~Templeton,
``Topologically Massive Gauge Theories,''
Annals Phys.\  {\bf 140}, 372 (1982)
[Annals Phys.\  {\bf 281}, 409 (2000)]
Erratum: [Annals Phys.\  {\bf 185}, 406 (1988)].


\bibitem{Banados:1992wn} 
M.~Banados, C.~Teitelboim and J.~Zanelli,
``The Black hole in three-dimensional space-time,''
Phys.\ Rev.\ Lett.\  {\bf 69}, 1849 (1992)
[hep-th/9204099].


\bibitem{Li:2008dq} 
W.~Li, W.~Song and A.~Strominger,
``Chiral Gravity in Three Dimensions,''
JHEP {\bf 0804}, 082 (2008)
[arXiv:0801.4566 [hep-th]].


\bibitem{Bergshoeff:2009hq} 
E.~A.~Bergshoeff, O.~Hohm and P.~K.~Townsend,
``Massive Gravity in Three Dimensions,''
Phys.\ Rev.\ Lett.\  {\bf 102}, 201301 (2009)
[arXiv:0901.1766 [hep-th]].


\bibitem{Bergshoeff:2014pca} 
E.~Bergshoeff, O.~Hohm, W.~Merbis, A.~J.~Routh and P.~K.~Townsend,
``Minimal Massive 3D Gravity,''
Class.\ Quant.\ Grav.\  {\bf 31}, 145008 (2014)
[arXiv:1404.2867 [hep-th]].


\bibitem{Clement:2009gq} 
G.~Clement,
``Warped AdS(3) black holes in new massive gravity,''
Class.\ Quant.\ Grav.\  {\bf 26}, 105015 (2009)
[arXiv:0902.4634 [hep-th]].


\bibitem{Setare:2015pva} 
M.~R.~Setare and H.~Adami,
``Entropy formula of black holes in minimal massive gravity and its application for BTZ black holes,''
Phys.\ Rev.\ D {\bf 91}, no. 10, 104039 (2015)
[arXiv:1501.00920 [hep-th]].


\bibitem{Cai:2016xho} 
  R.~G.~Cai, S.~M.~Ruan, S.~J.~Wang, R.~Q.~Yang and R.~H.~Peng,
  ``Action growth for AdS black holes,''
  JHEP {\bf 1609}, 161 (2016)
  [arXiv:1606.08307 [gr-qc]].



\bibitem{Brenna:2015pqa} 
  W.~G.~Brenna, R.~B.~Mann and M.~Park,
  ``Mass and Thermodynamic Volume in Lifshitz Spacetimes,''
  Phys.\ Rev.\ D {\bf 92}, no. 4, 044015 (2015)
  [arXiv:1505.06331 [hep-th]].
\bibitem{Carmi:2017jqz} 
  D.~Carmi, S.~Chapman, H.~Marrochio, R.~C.~Myers and S.~Sugishita,
  ``On the Time Dependence of Holographic Complexity,''
  JHEP {\bf 1711}, 188 (2017)
  [arXiv:1709.10184 [hep-th]].
\bibitem{Alishahiha:2019cib} 
  M.~Alishahiha, K.~Babaei Velni and M.~R.~Tanhayi,
  ``Complexity and Near Extremal Charged Black Branes,''
  arXiv:1901.00689 [hep-th].
\bibitem{Mezei:2016zxg} 
  M.~Mezei,
  ``On entanglement spreading from holography,''
  JHEP {\bf 1705}, 064 (2017)
  [arXiv:1612.00082 [hep-th]].
\bibitem{Tarrio:2011de} 
  J.~Tarrio and S.~Vandoren,
  ``Black holes and black branes in Lifshitz spacetimes,''
  JHEP {\bf 1109}, 017 (2011)
  [arXiv:1105.6335 [hep-th]].
\bibitem{An:2018xhv} 
  Y.~S.~An and R.~H.~Peng,
  ``Effect of the dilaton on holographic complexity growth,''
  Phys.\ Rev.\ D {\bf 97}, no. 6, 066022 (2018)
  [arXiv:1801.03638 [hep-th]].
  
\bibitem{Andrade:2013gsa} 
  T.~Andrade and B.~Withers,
  ``A simple holographic model of momentum relaxation,''
  JHEP {\bf 1405}, 101 (2014)
  [arXiv:1311.5157 [hep-th]].

  
\bibitem{Davison:2014lua} 
  R.~A.~Davison and B.~Goutéraux,
  ``Momentum dissipation and effective theories of coherent and incoherent transport,''
  JHEP {\bf 1501}, 039 (2015)
  [arXiv:1411.1062 [hep-th]].
  
\bibitem{Lehner:2016vdi} 
  L.~Lehner, R.~C.~Myers, E.~Poisson and R.~D.~Sorkin,
  ``Gravitational action with null boundaries,''
  Phys.\ Rev.\ D {\bf 94}, no. 8, 084046 (2016)
  [arXiv:1609.00207 [hep-th]].
 
\bibitem{Meng:2018vtl} 
  K.~Meng,
  ``Holographic complexity of Born–Infeld black holes,''
  Eur.\ Phys.\ J.\ C {\bf 79}, no. 12, 984 (2019)
  [arXiv:1810.02208 [hep-th]].

\bibitem{Cai:2017sjv} 
  R.~G.~Cai, M.~Sasaki and S.~J.~Wang,
  ''Action growth of charged black holes with a single horizon,''
  Phys.\ Rev.\ D {\bf 95}, no. 12, 124002 (2017)
  doi:10.1103/PhysRevD.95.124002
  [arXiv:1702.06766 [gr-qc]].

  
\end{thebibliography}
\end{document}